\documentclass[10pt, conference, compsocconf]{IEEEtran}
\usepackage{amsmath,graphicx}
\usepackage{float}
\begin{document}
\title{Basic Parallel and Distributed Computing Curriculum}


\author{\IEEEauthorblockN{Claude Tadonki}
\IEEEauthorblockA{Mines ParisTech - PSL Research University\\Centre de Recherche en Informatique (CRI) - Dept. Math\'ematiques et Syst\`emes\\
35, rue saint-honor\'e, 77305 Fontainebleau-Cedex (France)\\
claude.tadonki@mines-paristech.fr}
}

\maketitle

\begin{abstract} With the advent of multi-core processors and their fast
expansion, it is quite clear that {\em parallel computing} is now a genuine  
requirement in Computer Science and Engineering (and related) curriculum. In
addition to the pervasiveness of parallel computing devices, we should take into
account the fact that there are lot of existing softwares that are implemented
in the sequential mode, and thus need to be adapted for a parallel execution.
Therefore, it is required to the programmer to be able to design parallel
programs and also to have some skills in moving from a given sequential code to
the corresponding parallel code. In this paper, we present a basic educational
scenario on how to give a consistent and efficient background in parallel
computing to ordinary computer scientists and engineers.   

%The abstract goes here. DO NOT USE SPECIAL CHARACTERS, SYMBOLS, OR MATH IN YOUR
%TITLE OR ABSTRACT.

\end{abstract}


% \fi
\IEEEpeerreviewmaketitle

\section{Introduction}
In the past, {\em parallel computing} courses were dedicated to HPC specialists,
under appropriate prerequisites. This was due, on one hand, to the technical context, where standard processors were single-core, parallel computers being the corresponding clusters (shared or distributed memory). In addition, processors speed was increasing significantly (following the Moore's law), thus giving an argument to refrain from moving to parallel computing. Indeed, what one could achieve using a moderate cluster at a given time could be done a few years later using next generation processor. Therefore, as parallel computing could not be reasonably considered for basic issues, it was quite hard to motivate bringing it into standard courses. On the other hand, the basis to understand parallel computing and have hands on it show a significant gap from ordinary skills. Thus, one could understand a certain reluctance to such a heavy effort from both sides (students and teachers). Nowadays, the situation is no longer the same, and we have to bring parallel and distributed computing (at least at the basic level) into the standard. The corresponding courses have to be ready for a common audience.

{\em Parallel and Distributed Computing} (PDC) is a specialized topic, commonly encountered in the general context of {\em High Performance/Throughput Computing}. We mainly see three kind of material that could be considered when it comes to teaching PDC. 
\begin{itemize}
\item
First, the literature. There are numerous valuable books that cover general and/or specific aspects of PDC. General books that provide an overview of the topic with details on some selected aspects \cite{book:1, book:3, book:4, book:5, book:18}. Some books are more educational (tutorial approach with exercises and case studies) \cite{book:2, book:12, book:13, book:14}. Other manuals focus on specific architectures \cite{book:6, book:7}  or libraries (MPI, OpenMP, Pthreads) \cite{book:9, book:10}. There are books devoted to parallel algorithms design and  fundamental aspects (foundation, models, schedu\-ling, complexity) \cite{book:15, book:16, book:17, book:19, book:20, book:21}. 
\item
Second, conferences and assimilated events \cite{book:25, book:26, book:27, book:28, book:29, book:30} are good places to learn about advances and keep updated with new results. There are numerous events dedicated to PDC, some of them being tailored for student exposure and training through specialized tutorials.
\item
Third, summer/winter schools (or advanced schools) \cite{book:22, book:23, book:24} are good opportunities for specialized training, through an intensive few days course. Depending on the content, purpose, or audience, such schools are intended to develop a particular skill, or give a short but consistent PDC introductory course. For instance, an ordinary student who intends to do a PhD in parallel computing could use such schools as a starting point.    
\end{itemize}
Because of the prerequisites and a certain technical maturity needed to deal with parallel computing, we think that PDC courses could be reasonably considered at nearly the end of the undergraduate curriculum. At his level, the aim could be to have the students being able to design an intermediate level parallel program. The courses could be organized around that objective, taking into account the background of the student and what is really needed at that point. A typical scenario could include:
\begin{itemize}
 \item  general introduction
 \item  parallel computation models
 \item  distributed memory paradigm
 \item  shared memory paradigm
 \item  instruction level parallelism
 \item  performance evaluation
 \item  debugging and profiling tools
 \item  virtualization and simulators
 \item  specialized frameworks
\end{itemize}
In addition to how to connect the selected chapters, it is important to teach them at the right and appropriate level. Indeed, as we target an undergraduate curriculum, and following the aforementioned global objective, we just need to stay in the necessary scope. In addition, the way each chapter is introduced and handled is important. We classify the selected chapters into three main group and develop our argumentation accordingly. The following section provides a global overview of the PDC course. Next, section \ref{chap2} describes how to provide introductory PDC elements. In section \ref{chap3}, we expose different ways to implement parallelism. We discuss about parallel programs execution in section \ref{chap4}. Section \ref{chap5} concludes the paper.

\section{Course overview}\label{chap1}
\begin{figure}[h]
\centering
\includegraphics[scale=0.6]{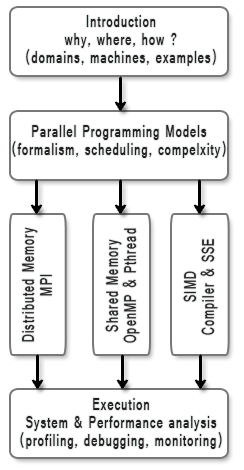}
\caption{Integrated overview of the PDC course}
\label{fig1}
\end{figure}

A basic scenario for a consistent PDC course is displayed in figure \ref{fig1}. 

The first thing to do is to buy the attention of the students by providing some illustrative examples and outline some convincing motivation items. Another purpose with the examples is to help understanding why it is important (sometimes vital) to compute faster. As the need for speed and the programming model are  different from one application to another, this could be the time to talk about domain classification. Next, comes the hardware aspect, means {\em parallel machines} and accelerators. For undergraduate students, the topic of the {\em accelerators} \cite{nrj2, tad1} could be postponed for postgraduate level or left as an option for particularly motivated students. This part could be ended by discussing each of the dissemination media listed in the introduction section. After this course unit, students are ready to enter into the subject. Energy concerns \cite{nrj, tad4, nrj2} could be mentioned too.
\section{Basic of parallelism}\label{chap2}
As parallel computing means simultaneous processing of several tasks, it is important here to introduce the notion of {\em dependence} \cite{karp, tad2,depp}. Dependences are the cause of {\em synchronization}, {\em data communication}, and sub-optimal performances. Students should be able to identify the main dependences between tasks and appreciate the potential of parallelism related to a given application. Deriving a (good) parallel scheduling could be the next point. Scheduling algorithms and methodologies could be explained together with their associated formalism (tasks graph, recurrence equations, ...). Some basic elements of performance prediction could be presented here, leaving the aspect of {\em pure complexity} for specialized students. Once the students are familial with the concept of parallelism, it could be time to consider the implementation aspect.
\section{Ways to implement parallelism}\label{chap3}
What could be done here is to present different level of parallelism and then focus on the most commonly considered solutions, namely MPI for the message passing paradigm, OpenMP and Pthread for thread level parallelism, and SSE for instruction level parallelism \cite{book:6, tad6, book:11}. It is not necessary here to to into deeper details on each programming model. It is rather important to have the students being able to derive effective implementations for some basic examples and understand that achieving a high speed program could come from a hybrid implementation. In general, there is a software gap between the hardware potential and the performance that can be attained by practical programs. Thus, it is important to handle the execution correctly and understand the performance point.
\section{Running time}\label{chap4}
Compiling and running a parallel program is the last point of our PDC course scenario. After having the program running, it is important to know how to measure its performance, and thus appreciate the impact of the implemented parallelism. As for sequential programs, some mistake could have been done, either at the design stage or at the programming stage. Teaching the use of debugging techniques and tools could be considered, with the aim of being able to detect and fix programming mistakes or system issues. Hardware issues could be mentioned but not covered. 
\section{Discussions and Conclusion}\label{chap5}
Teaching parallel and distributed programming at any level is a genuine requirement nowadays. In order to fulfill this crucial need, PDC course should be incorporated into standard scientific and engineer curriculum. There is certainly a pedagogical effort to bring this topic, previously reserved for specialists, into the standard. Whenever possible, the earlier it is done, the better. In this paper, we propose consistent scenario that could apply at the undergraduate level. We also thing that using well designed simulators could be very useful for this training task.

\end{document}